\documentclass[aps,prl,showpacs,twocolumn,amsmath,amssymb,superscriptaddress,citeautoscript]{revtex4}

\usepackage{graphicx}
\usepackage{dcolumn}
\usepackage{bm}
\usepackage[caption=false]{subfig}
\usepackage{amssymb}
\usepackage{amsmath}
\usepackage{commath}
\usepackage{graphicx}
\usepackage{verbatim}
\usepackage{hyperref}

\hypersetup{
    colorlinks=true, 
    citecolor=blue,
    linkcolor=blue, 
    urlcolor=red, 
    linktoc=all 
}

\begin{document}

\preprint{APS/123-QED}

\title{Large effects of subtle electronic correlations on the energetics of vacancies in $\alpha$-Fe}

\author{Pascal Delange}
\affiliation{ Centre de Physique Th\'eorique, \'Ecole Polytechnique, CNRS, Universit\'e Paris-Saclay, 91128 Palaiseau Cedex, France }
\author{Thomas Ayral}
\affiliation{ Institut de Physique Th\'eorique (IPhT), CEA, CNRS, UMR 3681, 91191 Gif-sur-Yvette, France }
\author{Sergei I. Simak}
\affiliation{ Department of Physics, Chemistry and Biology (IFM), Link\"{o}ping University, SE-58183 Link\"{o}ping, Sweden }
\author{Michel Ferrero}
\affiliation{ Centre de Physique Th\'eorique, \'Ecole Polytechnique, CNRS, Universit\'e Paris-Saclay, 91128 Palaiseau Cedex, France }
\affiliation{Coll\`{e}ge de France, 11 place Marcelin Berthelot, 75005 Paris, France}
\author{Olivier Parcollet}
\affiliation{ Institut de Physique Th\'eorique (IPhT), CEA, CNRS, UMR 3681, 91191 Gif-sur-Yvette, France }
\author{Silke Biermann}
\affiliation{ Centre de Physique Th\'eorique, \'Ecole Polytechnique, CNRS, Universit\'e Paris-Saclay, 91128 Palaiseau Cedex, France }
\affiliation{Coll\`{e}ge de France, 11 place Marcelin Berthelot, 75005 Paris, France}
\author{Leonid Pourovskii}
\affiliation{ Centre de Physique Th\'eorique, \'Ecole Polytechnique, CNRS, Universit\'e Paris-Saclay, 91128 Palaiseau Cedex, France }
\affiliation{Coll\`{e}ge de France, 11 place Marcelin Berthelot, 75005 Paris, France}

\date{\today}

\begin{abstract}   
We study the effect of electronic Coulomb correlations on the vacancy formation energy in paramagnetic $\alpha$-Fe within \emph{ab initio} dynamical mean-field theory. The calculated value for the formation energy is substantially lower than in standard density-functional calculations and in excellent agreement with experiment. The reduction is caused by an enhancement of electronic correlations at the nearest neighbors of the vacancy. This effect is explained by subtle changes in the corresponding spectral function of the $d$-electrons. The local lattice relaxations around the vacancy are  substantially increased by many-body effects.

\end{abstract}

\maketitle



Point defects, such as vacancies, play an important role for the mechanical and thermodynamic properties of materials\cite{callister2007materials}. However, the experimental determination of vacancy formation or migration energies is difficult. Even the best available techniques, the differential dilatometry and the positron annihilation spectroscopy, suffer from large error bars, and the discrepancies between different measurements on one and the same material may be significant. 
Therefore, \emph{ab initio} theoretical calculations are an indispensable tool for developing a better understanding of the defect properties of materials\cite{kresse_rmp_2014}.

Early density functional theory (DFT) calculations in the local density approximation (LDA) have predicted formation energies of vacancies in simple metals in good agreement with experiment\cite{nieminen_simple_metals_1977,picket_Al_vac_1981}. Despite a large body of successful calculations, it has later been recognized that the nice agreement with experiment could often be the effect of the cancellation of errors in the exchange and correlation parts of the density functional \cite{carling2000}. As has been discussed by Ruban \cite{ruban2016}, despite the structural simplicity of vacancies, their energetics is still one of the least reliable physical properties determined in first-principles calculations.

In transition metals, where the open $d$ shells are often poorly described in LDA or  the generalized gradient approximation (GGA), the quality of results of DFT calculations for point defect properties is rather unpredictable and strongly material-dependent. There have been several attempts to improve the available functionals (see, e.g. Refs. \cite{armiento2005,AM2008,nazarov2012,grabowsky2014}). We notice that the predicted vacancy formation energies seem to be especially poor for 3$d$ transition metals, for which many-body effects are fairly important, in particular in the paramagnetic state and body-cented cubic (bcc) crystal structure\cite{korhonen1995vacancy}. 
Likewise, DFT has limitations for point defect calculations in correlated lanthanide or actinide oxides with $4f$ or $5f$ electrons, for example in the case of uranium oxides used in the nuclear industry\cite{amadon_UO2_2013}. 


Among the 3$d$ transition metals, iron is a particularly complex system, where the strength of electronic correlations is very sensitive to the lattice structure and magnetic state. However, from a practical point of view, vacancies in iron and steels are of particular interest because they affect a number of important characteristics of the metal, e.g. toughness and ductility. Iron's low-temperature ferromagnetic bcc $\alpha$ phase is a weakly-renormalized Fermi liquid\cite{pourovskii2014} well described within DFT \cite{Moruzzi1986,Stixrude1994,Soderlind1996}. However, the same $\alpha$-Fe in the high-temperature paramagnetic phase exhibits a strongly-correlated non-Fermi-liquid behavior \cite{Katanin2010,leonov2011,pourovskii2013} with  DFT calculations failing to describe its structural parameters (lattice constants, bulk modulus, or even the shape of the crystal)  \cite{leonov2011}.  The low-temperature paramagnetic hexagonal $\epsilon$ phase stabilized by pressure is also rather strongly correlated \cite{pourovskii2014} (though less so as compared to paramagnetic $\alpha$-Fe), exhibiting a large electron-electron scattering contribution to the resistivity \cite{Jaccard2005} as well as unconventional superconductivity \cite{Shimizu2000,Mazin2002}. All this hints at a strong sensitivity of many-body effects in Fe to local disturbances of the crystalline order (e.g., to point defects) that cannot be captured within standard DFT. Indeed, extensive DFT calculations of $\alpha$-Fe \cite{becquart_iron_2001,soderlind_vac_bcc_metal_2000,mizuno2001first,connetable_multivac_iron_2013} consistently predict a monovacancy formation energy about 30 to 40 \% higher than the measured value. In contrast to other 3$d$ metals, this formation energy has been somewhat reliably determined thanks to extensive experiments \cite{schaefer_postitron_1977,kim_positron_1978,Matter_positron_1979,schepper_positron_1983,schaefer_positron_1987}.

The deficiencies of standard DFT to describe $\epsilon$-Fe and paramagnetic $\alpha$-Fe have been successfully corrected by combining it with a dynamical mean-field theory (DMFT) \cite{georges_dmft_1996, vollkot}  treatment of the local repulsion between 3$d$ electrons. {\it Ab initio} calculations  using this DFT+DMFT approach \cite{anisimov_1997,lichtenstein_LDA_DMFT} were able to reproduce the ground-state properties and phonon spectra of the $\alpha$-phase \cite{leonov2011,Leonov2012} as well as the equation of state of $\epsilon$-Fe \cite{pourovskii2014}. It is thus likely that an explicit treatment of many-body effects within DMFT will also correct the severe problems of DFT in describing point defects in iron. 

Hence, in the present work we have developed the state-of-the-art DFT+DMFT method  \cite{kotliar_elec_struc_dmft_2006, biermann_ldadmft, LechermannPRB74-06, AnisimovPRB71-05} into a scheme for studying vacancy properties. We have applied our technique to a single vacancy in $\alpha$-Fe, computing the electronic structure around the vacancy as well as the vacancy's formation energy, taking into account local lattice distortions around the defect. We do not treat here the high-temperature face-centered cubic (fcc) phase, nor temperatures close to the melting point, where the influence of the (anharmonic) lattice vibrations may play a crucial role \cite{carling2000,andersson2004,grabowsky2014} and the bcc phase is stabilized again. Compared to DFT-LDA, a significant reduction of the theoretical formation energy is obtained, with calculated values in remarkable agreement with experimental estimates \cite{schaefer_postitron_1977,kim_positron_1978,Matter_positron_1979,schepper_positron_1983,schaefer_positron_1987}. We trace back this reduction to rather subtle effects of the vacancy on the local density of states and hybridization with its nearest neighbors.


\begin{figure}
\subfloat[2x2x2 supercell\label{sfig:2x2x2_cell}]{
  \includegraphics[width=41mm]{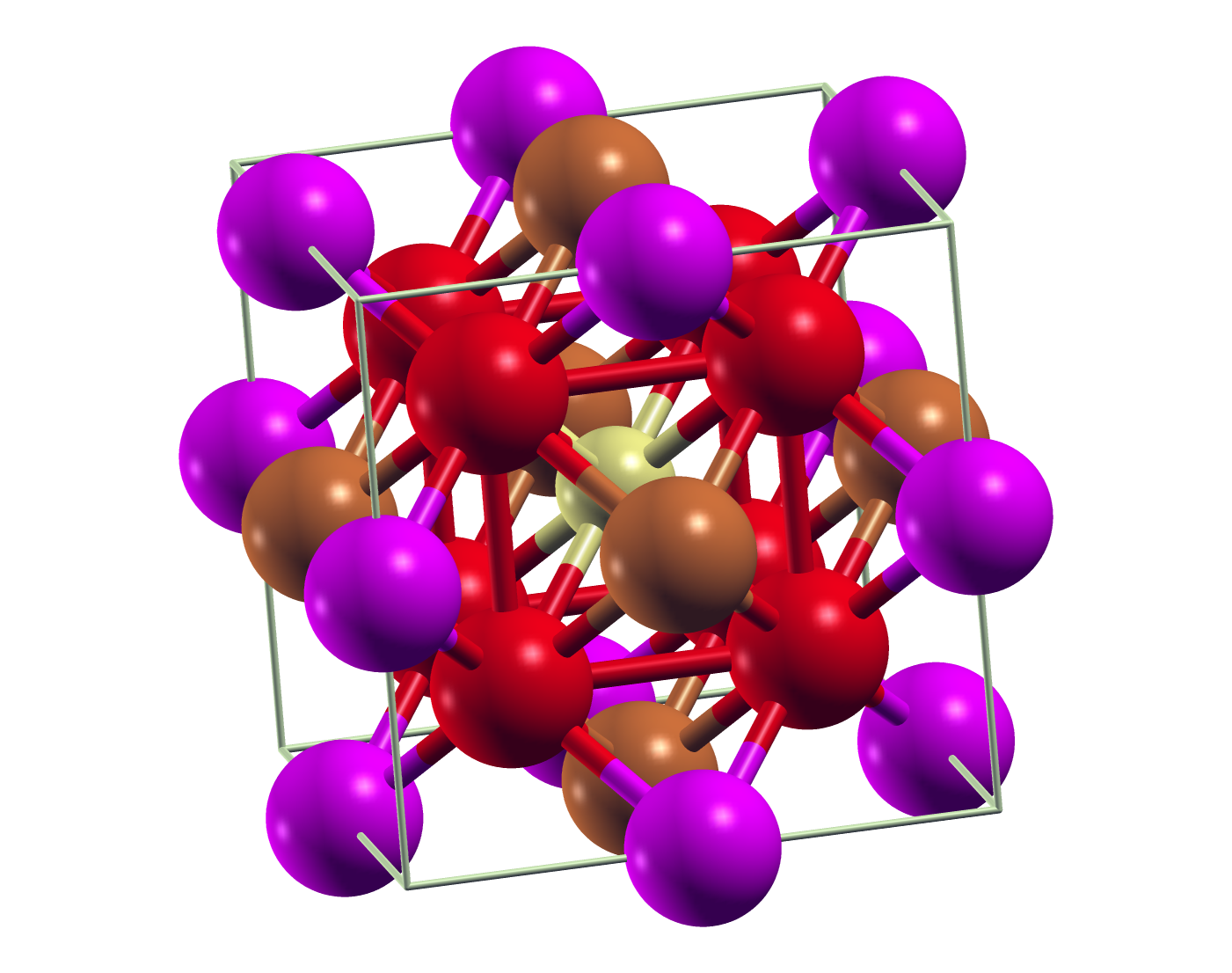}
}  \hfill
\subfloat[3x3x3 supercell\label{sfig:3x3x3_cell}]{
  \includegraphics[width=41mm]{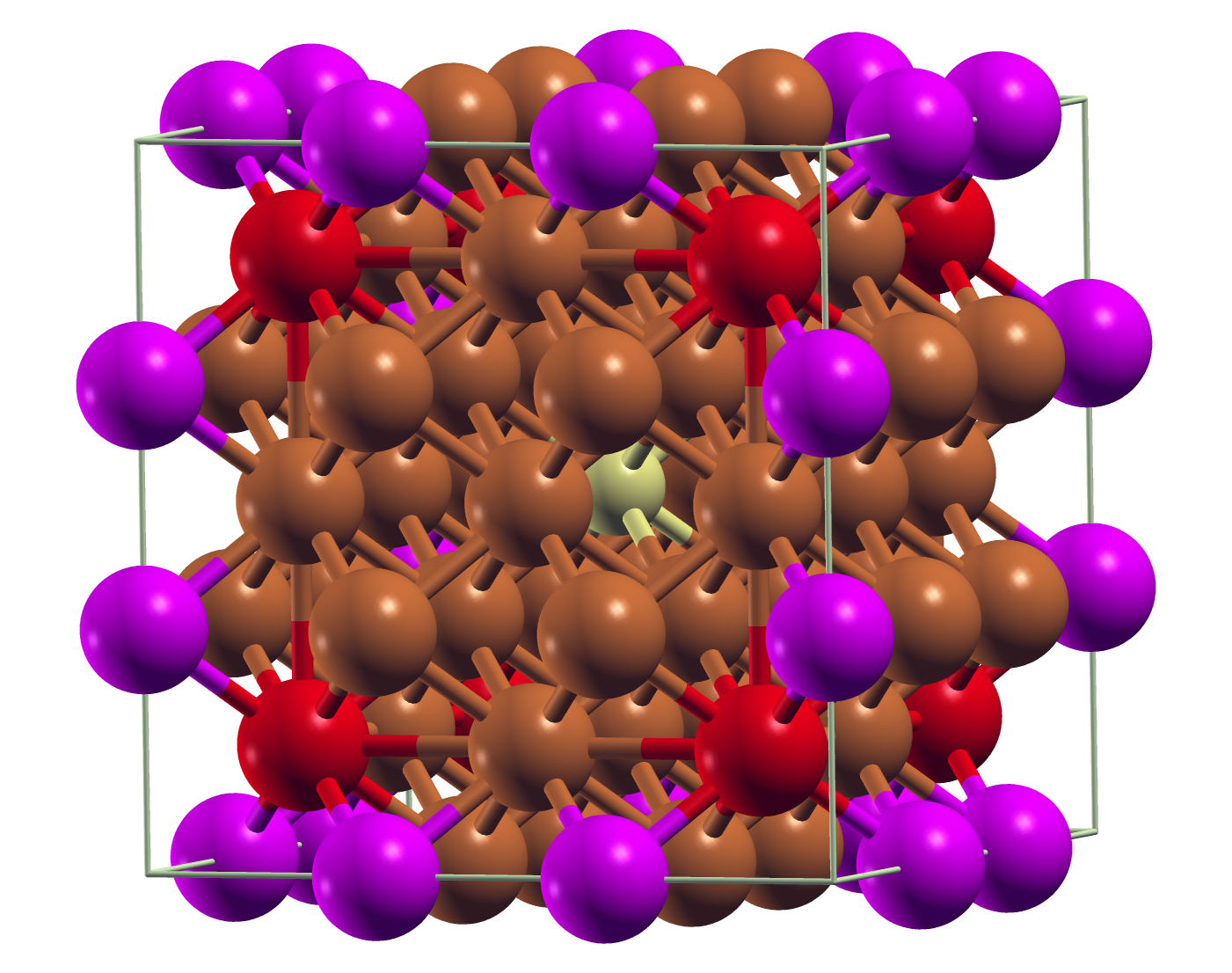} 
}
\caption{(Color online) The $2 \times 2 \times 2$ and $3 \times 3 \times 3$  supercells with the vacancy in the corner. Different colors indicate the atom nearest to the vacancy (red), the second nearest (purple), and the furthest (the central atom, yellow).}
\label{fig:supercell}
\end{figure}

We model a single vacancy in bcc Fe using the $2\times 2\times 2$ and $3 \times 3 \times 3$ cubic supercells represented in Fig.~\ref{fig:supercell}, with the vacancy placed at the origin of the supercells. We compute the vacancy formation energy from the supercell total energy using the standard formula 

\begin{equation}
E^f_{\mathrm{vac}}=E^{\mathrm{vac}}(N-1) - \frac{N-1}{N}E^{\mathrm{no\ vac}}(N),
\end{equation}
\label{eq:formation_energy}
where $N$ is the number of atoms in the ideal supercell, $E^{\mathrm{no\ vac}}(N)$ is the total energy of the ideal supercell containing $N$ atoms and no vacancy, and $E^{\mathrm{vac}}(N-1)$ is the total energy of the same supercell with a vacancy (hence $N-1$ atoms). $N$ is 16 in the 2x2x2 supercell and 54 in the 3x3x3 supercell, corresponding to vacancy concentrations of respectively $6.25$ and $1.85\%$, respectively.


Our calculations have been carried out using a fully charge self-consistent implementation of  DFT+DMFT\cite{Aichhorn2009,Aichhorn2011} based on the TRIQS package \cite{triqs_main,triqs_dft_tools}. This implementation  is based on the full potential linearized augmented plane-wave (FLAPW) Wien2k code \cite{blaha2001wien2k}. 
The on-site density-density interaction between those orbitals is parametrized by the Slater parameter $F_0=U=4.3$~eV and the Hund's rule coupling $J=1.0$~eV that were previously used in the DFT+DMFT calculations of $\alpha$ and $\epsilon$-Fe of Ref.~\cite{pourovskii2014}. The calculation of a vacancy formation energy using supercells with seven inequivalent atomic sites -- which, to the best of our knowledge, has hitherto never been achieved -- has become possible thanks to the use of a continuous-time quantum Monte-Carlo (CTQMC) hybridization expansion algorithm \cite{Gull2011} in the segment representation for the solution of the local impurity problems (Lechermann et al. have reported about the effect of vacancies on magnetism within DFT+DMFT \cite{lechermann_vac_dmft_2016,lechermann_vac_dmft_2015}).
All DFT+DMFT calculations were performed at a temperature of $1162\text{K}$ in the paramagnetic phase of $\alpha$-Fe, at its experimental lattice parameter of 2.86~\AA. The experimental volume of $\alpha$-Fe was almost exactly reproduced in Ref.~\cite{pourovskii2014} with identical calculation parameters.



\begin{figure}
        \centering
        \includegraphics[width=85mm]{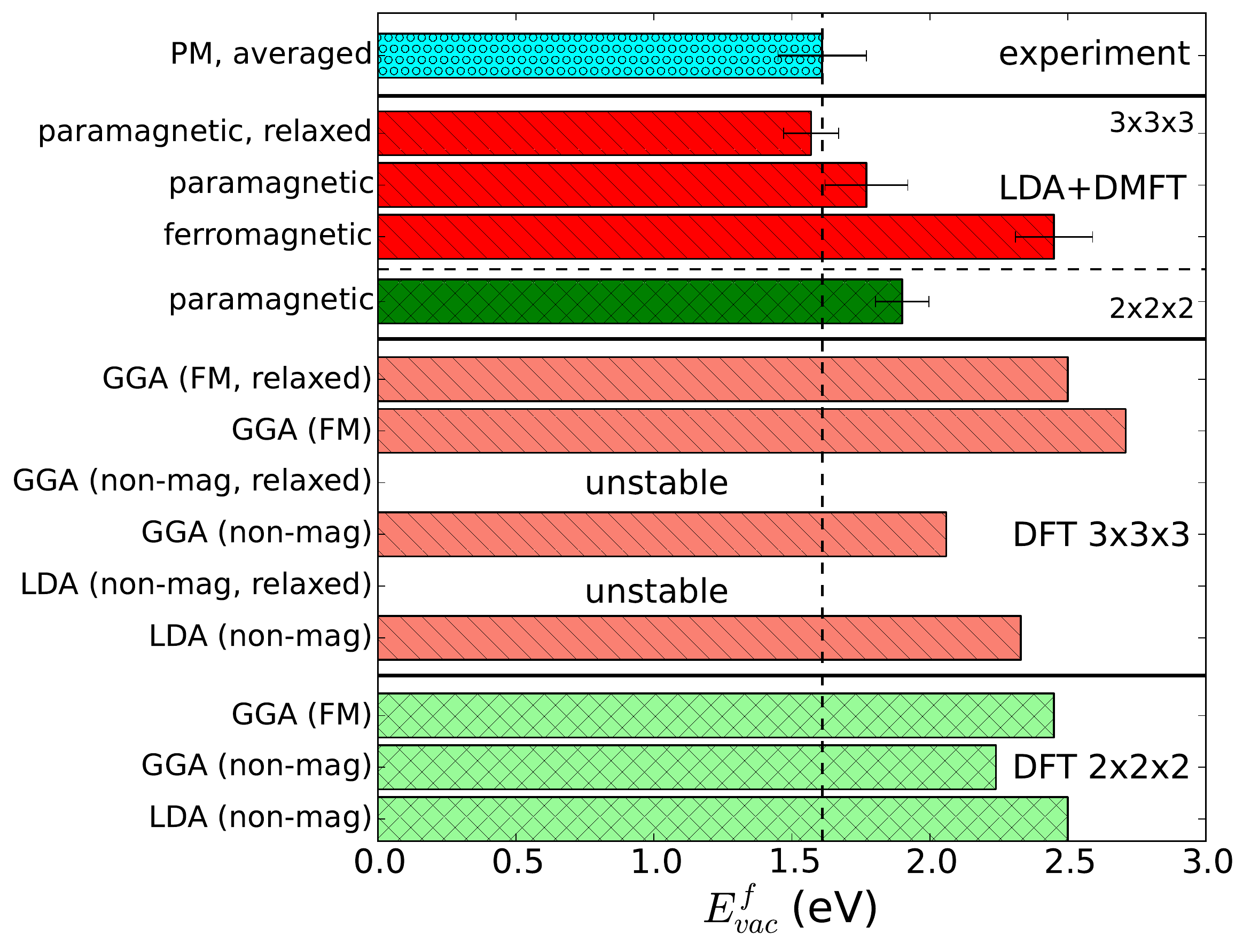}
        \caption{Vacancy formation energies calculated by different methods (LDA, GGA and DFT+DMFT) in the different setups: small and large supercell, relaxed or not. The average of experimental values is shown for comparison\cite{schaefer_postitron_1977,kim_positron_1978,Matter_positron_1979,schepper_positron_1983,schaefer_positron_1987} (Color online)}
        \label{fig:energies}
\end{figure}

The vacancy formation energies obtained within DFT+DMFT together with different DFT results and experimental values are shown in Fig.~ \ref{fig:energies} (see also Table I of the Supplementary Material~\cite{supp_tab}). The resulting value for $E^f_\mathrm{vac}$ in DFT+DMFT  is 1.77~eV for the unrelaxed 54-atom supercell with lattice relaxations reducing it further to $E^f_\mathrm{vac} = 1.56\pm 0.13$eV, in excellent agreement with the mean experimental value of about 1.6~eV. We also calculated $E^f_\mathrm{vac}$ within DFT+DMFT for the unrelaxed ferromagnetic phase obtaining a higher value of 2.45$\pm0.15$eV.
 Experiments indeed seem to confirm that $E^f_\mathrm{vac}$ in the ferromagnetic phase should be larger than in the nonmagnetic one\cite{schepper_positron_1983,schaefer_positron_1987}, though direct low-temperature measurements of $E^f_\mathrm{vac}$ in the ferromagnetic phase with positron annihilation spectroscopy are difficult  and all reported values have been extrapolated from high-temperature measurements.

DFT (GGA) calculations assuming ferromagnetic bcc Fe predict a significantly larger value $E^f_\mathrm{vac}$  of 2.70 and 2.50~eV for an unrelaxed and a fully relaxed cell, respectively. Hence, one sees that many-body effects included within DMFT reduce $E^f_\mathrm{vac}$ for the paramagnetic phase by about 0.9~eV. The impact of correlation effects for ferromagnetic $\alpha$-Fe is less significant, in agreement with the predicted suppression of dynamic correlations in this phase\cite{pourovskii2014}.
The vacancy formation energies obtained with nonmagnetic LDA or GGA calculations are closer to the measured values, with $E^f_\mathrm{vac} \approx 2.05$eV in GGA. They have, however, a limited physical meaning: DFT in general fails dramatically for the paramagnetic phase, which is reflected by the fact that $\alpha-$Fe is not dynamically stable and the predicted lattice parameter would be significantly smaller in nonmagnetic LDA or GGA. Hence, using our relaxed positions in a nonmagnetic DFT calculation gives an (absurd) negative vacancy formation energy.  Thus, the somewhat reduced value of $E^f_\mathrm{vac}$ in nonmagnetic DFT calculations compared to ferromagnetic ones may be due to a spurious cancellation of errors. 

\begin{figure}
        \centering
        \includegraphics[width=85mm]{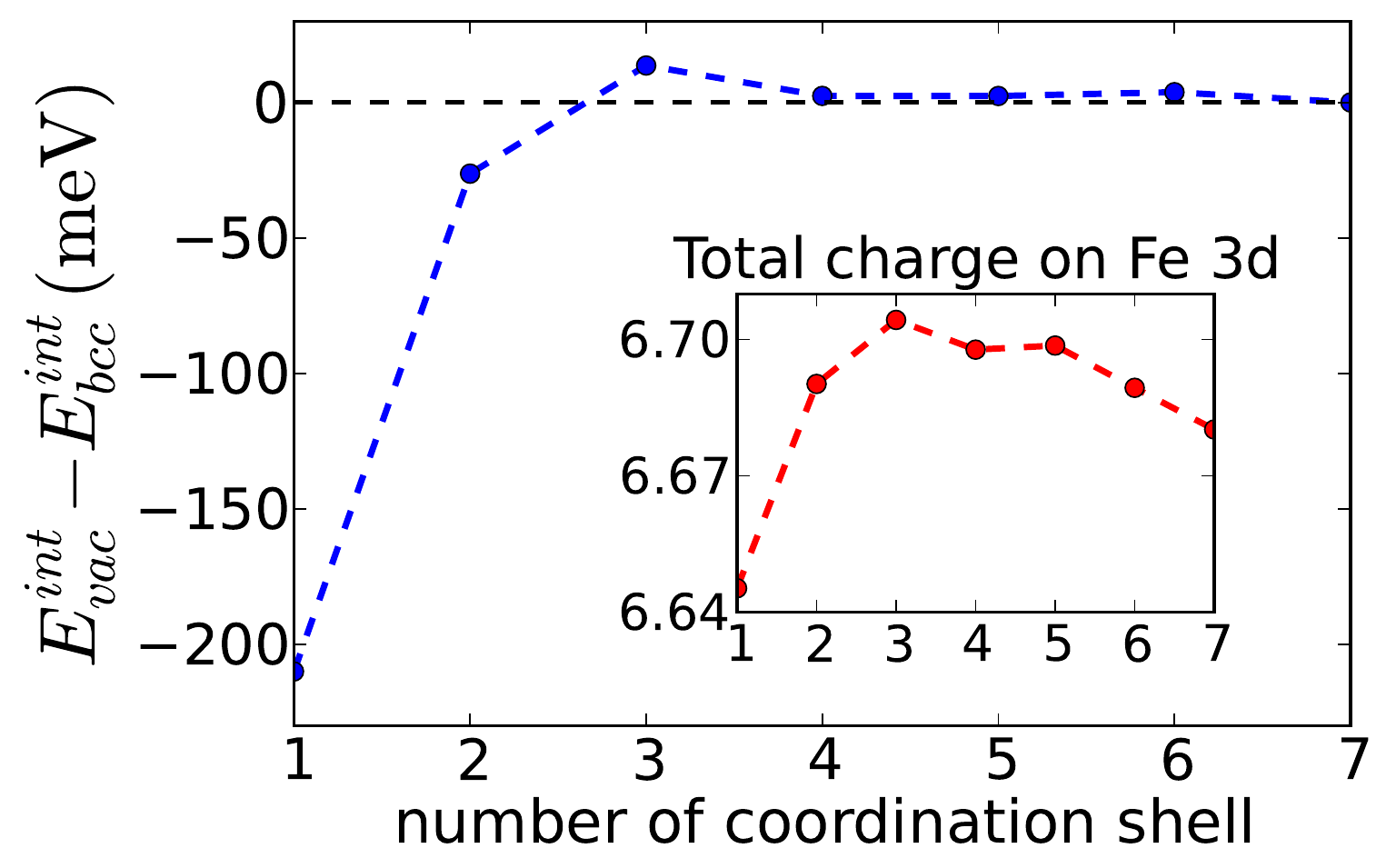}
        \caption{Difference in the interaction energy per atom, before and after adding a vacancy. Inset : Fe 3$d$ charge in the cell with vacancy.}
        \label{fig:int_energy}
\end{figure}

The total energy in DFT+DMFT is
\begin{equation}
E^{\mathrm{int}}_{\mathrm{DMFT}} = \mathrm{Tr}(\hat{\epsilon}_k\hat{\rho}_k^{\mathrm{DMFT}}) + E[\rho^{\mathrm{DMFT}}] + (E_{\mathrm{Hub}} - E_{\mathrm{DC}})
\label{eq:E_tot}
\end{equation}
where $\hat{\rho}_k^{\mathrm{DMFT}}$ is the density matrix for crystal momentum $k$, $\hat{\epsilon}_k$ the corresponding LDA hamiltonian and $E[\rho^{\mathrm{DMFT}}]$ only depends explicitly on the charge density. $E_{\mathrm{Hub}} =  \frac{1}{2}\sum_{ij}U_{ij}\left<n_in_j\right>  $ is the Coulomb interaction between Fe $3d$ electrons ($i$ and $j$ are orbital indices and $U_{ij}$ is the density-density Coulomb matrix), and $E_{\mathrm{DC}}$ is the double-counting term that estimates the energy already present in LDA (see Supplementary Material for the details~\cite{supp_tab}). When one removes an atom from the cell to create a vacancy, all three terms in (\ref{eq:E_tot}) change. Figure \ref{fig:int_energy} shows the difference in the third term, $E^{int} = E_{\mathrm{Hub}} - E_{\mathrm{DC}}$, on each respective atom of the supercell before and after removing an atom. Summing this up and taking into account the multiplicity of the atoms in the cell yields a change $\Delta E^{\mathrm{int}} \approx$ -1.6~eV, that is compensated by a larger change in $E[\rho^{\mathrm{DMFT}}]$ due to a redistribution of the charge density, as wavefunctions from DFT+DMFT are more localized. The contributions from the second and third coordination shells compensate one another, so that the net change in the interaction energy only comes from the first nearest neighbor. This is due to good metallic screening, and is in good agreement with embedded atom method calculations of iron vacancies near a surface\cite{wang_iron_vac_surface_2010} that show the vacancy formation energy to become equal to the bulk value for the vacancy located in  the third layer or deeper.

\begin{figure}
        \centering
        \includegraphics[width=85mm]{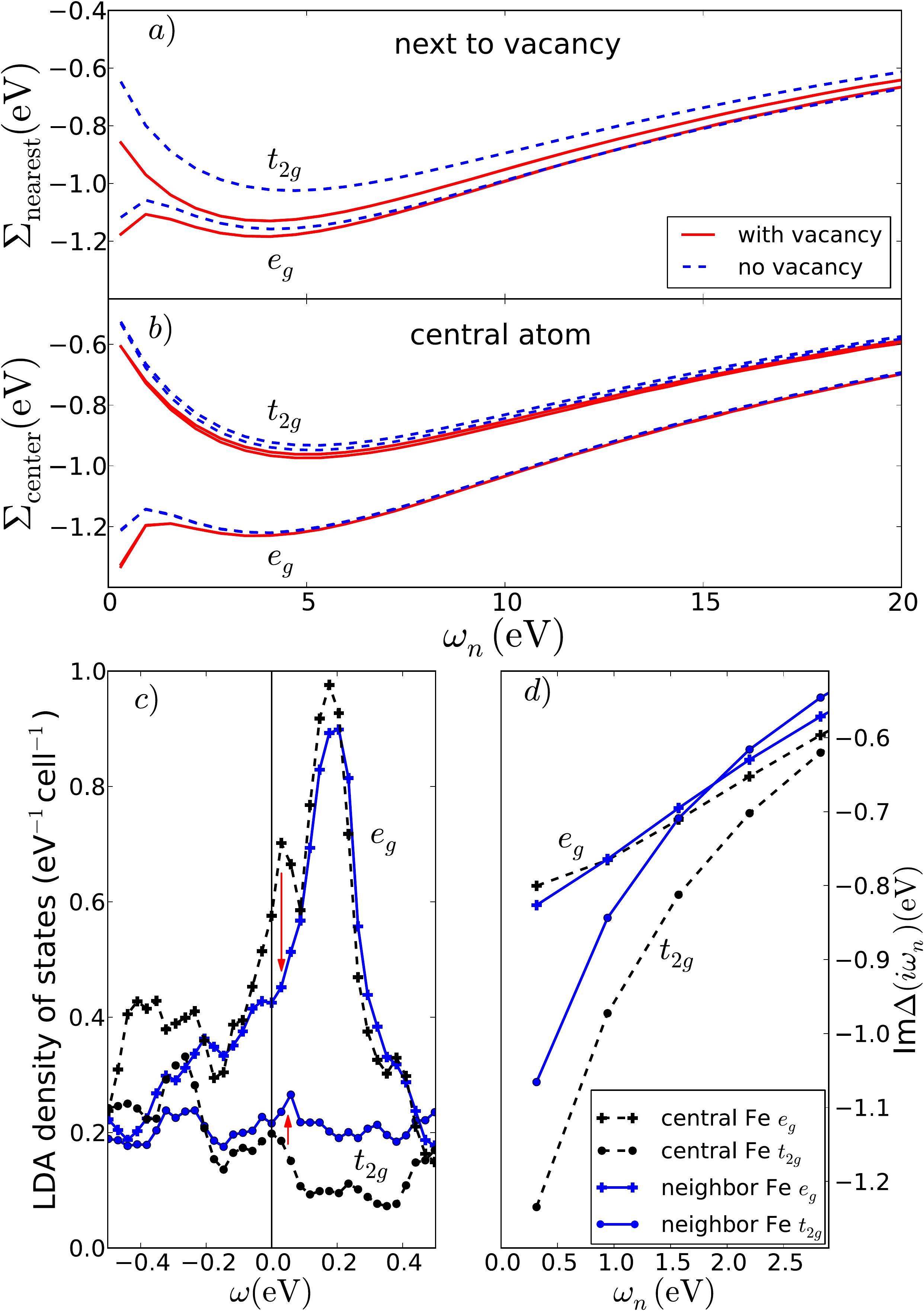}
        \caption{Imaginary part of the Matsubara self-energies for (a) the vacancy nearest neighbor and (b) central atom in the 3x3x3 supercell with a vacancy present (red, full) or without it (blue, dashed). Correlations become stronger on the atom nearest to the vacancy while the difference between $e_g$ and $t_{2g}$ is strongly reduced. c) LDA density of states around the Fermi level and d) Hybridization function on the Matsubara axis for the nearest neighbors (blue, full) to the vacancy, and for the central atom (black, dashed). The full Fe 3$d$ DOS is shown in the Supplementary Material~\cite{supp_tab}
         (Color online)}
        \label{fig:self-energies}
\end{figure}

The self-energy of the vacancy's first coordination shell shows a significant difference from the bulk bcc-Fe self-energy, as shown in Fig.~\ref{fig:self-energies}ab. $t_{2g}$ states, but also $e_g$ states to a lesser extent, become more strongly correlated (less coherent) with a larger Im$\Sigma(i\omega)$. This difference almost vanishes for the self-energy of the atoms further than the nearest neighbor, in agreement with the variation of the interaction energy shown in Fig.~\ref{fig:int_energy}. Stronger correlations on the atoms near the vacancy imply that a more correct description of the $3d$ electrons of the Fe atoms in DFT+DMFT, already important to predict the crystal structure and lattice parameter, is especially crucial when estimating the energetics of the vacancy and indeed leads to a smaller formation energy. Note that the self-energies are slightly atom-dependent even in the absence of a vacancy in our calculations, due to an artificial symmetry-breaking in the supercell in DFT calculations and the non-rotational invariance of the density-density Hubbard hamiltonian. However, we compare self-energies and interaction energies in a consistent, atom-to-atom way.

The enhancement of the nearest-neighbor self-energy can be traced back to a change in the hybridization function. As can be seen in Figure \ref{fig:self-energies}d, the imaginary-frequency hybridization function, in particular for the $t_{2g}$ states,  is reduced at low frequencies for the atom near the vacancy. This reduction is due to an increase in the corresponding $t_{2g}$ partial density of states (DOS) in the vicinity of the Fermi level, $E_F$, as one can see in Fig.~ \ref{fig:self-energies}c. A larger DOS at $E_F$ induces a suppression of  low-energy hopping leading to stronger correlation\cite{Mravlje2011,pourovskii2013}: at the first iteration of DMFT, Im$\Delta(i0^+)=-\pi \rho_F/\left[ \textrm{Re}G_{loc}(i0^+)^2 + (\pi\rho_F)^2 \right] \approx -1/(\pi\rho_F) $, with $\rho_F$ the LDA DOS. The enhancement of the nearest-neighbor $e_g$ self-energy is smaller and the corresponding DOS at $E_F$ even decreases compared to the bulk case. This decrease in the value the of DOS exactly at $E_F$ is compensated by an overall narrowing of the $e_g$ peak in the vicinity of $E_F$, see Fig.~ \ref{fig:self-energies}c. Hence, the resulting hybridization function for $e_g$ is still suppressed starting from the second Matsubara point.

\begin{figure}
 \includegraphics[width=85mm]{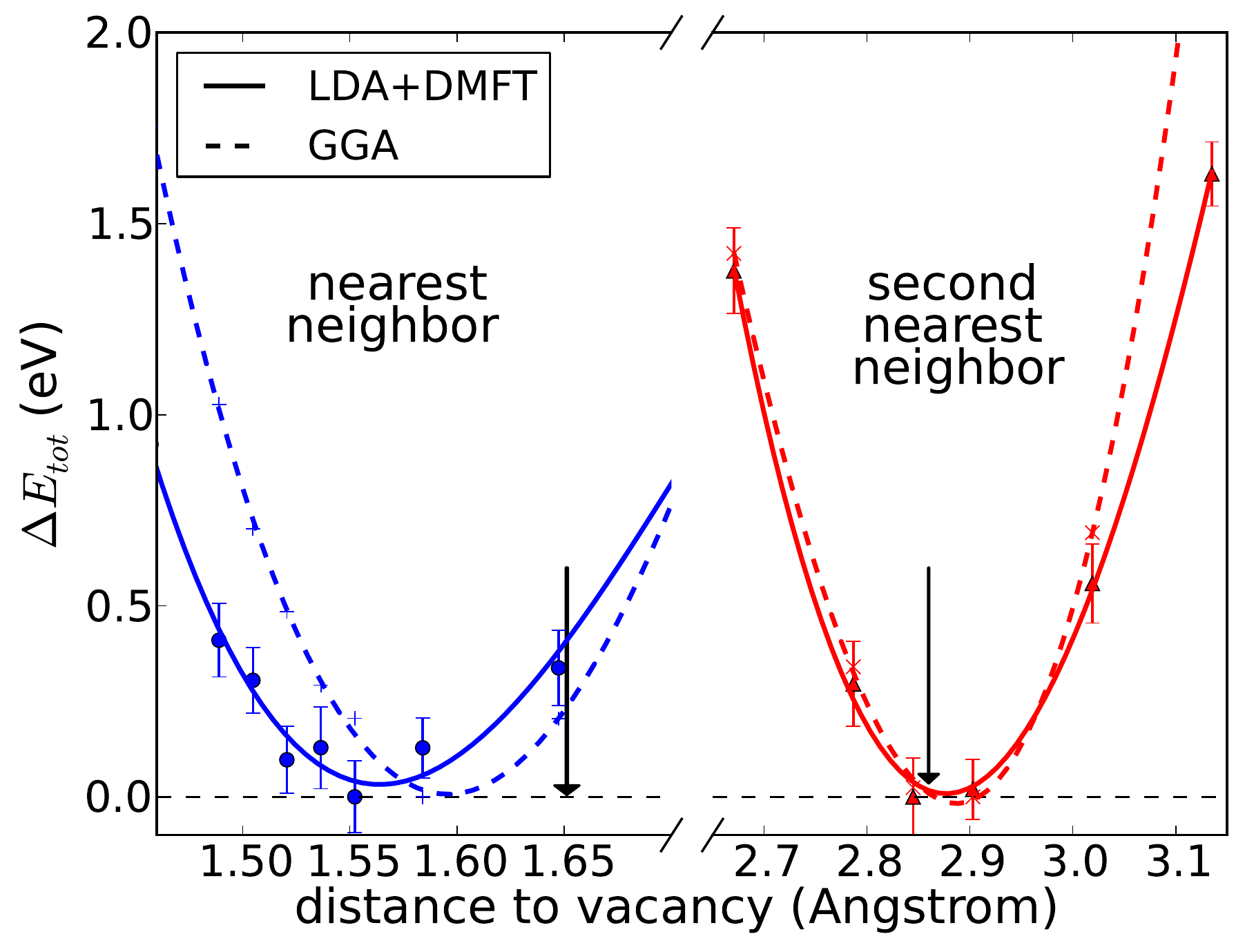} 
\caption{Total energy vs distance to the vacancy for the first nearest neighbor (blue) and the second nearest neighbor (red), in DFT+DMFT (full line) and GGA (dashed line). The black arrows show the position of the atoms in the unrelaxed bcc supercell. (Colors online)}
\label{fig:relaxation}
\end{figure}

We performed a relaxation of the atoms around the vacancy in two steps, in order to reduce the computational effort. Indeed, to consistently calculate atomic forces in DFT+DMFT it is not sufficient to calculate them in the DFT part of the scheme\cite{leonov_forces2014}. We first performed the full relaxation in spin-polarized GGA (at its corresponding theoretical volume) obtaining a shift of the first coordination shell towards the vacancy by about 4\%, and a shift of the second coordination shell away from the vacancy by about 1.5\%. All the other atoms do not move significantly. These GGA-relaxed positions are in agreement with previous calculations\cite{becquart_iron_2001}. Then the positions of the two first nearest neighbors were relaxed within DFT+DMFT. In Fig.~\ref{fig:relaxation} we show the total energy (minus an offset depending on the method used, GGA or DFT+DMFT) of the supercell as a function of the relaxed position of the nearest and second nearest neighbor of the vacancy. Each site was moved independently, preserving the symmetry of the cell, while the positions of others were fixed at their fractional  GGA values. We obtain the following results: The first nearest neighbor relaxes by another 1.8\% towards the vacancy, for a total relaxation of 5.7\% inwards changing the total energy by about 0.1eV. One sees that many-body effects have a significant impact on the nearest-neighbor relaxation, enhancing it by almost 50\%. Meanwhile, the second nearest neighbor relaxes back towards the vacancy by 0.8\%, for a total relaxation of 0.7\% outwards, with a negligible change in the total energy when compared to the GGA relaxed position. Overall, relaxing the two first coordination shells in full DFT+DMFT reduces the vacancy formation energy by 0.21~eV.

In conclusion, we have shown that local many-body effects are crucial to explaining a relatively low vacancy formation energy in $\alpha$-Fe. The presence of a vacancy induces rather subtle changes in the local electronic structure of its surroundings,  leading to a moderate increase in the strength of correlations at neighboring sites. This moderate increase has, however, a very significant impact on the vacancy energetics. When the effect of local relaxations is included, the calculated vacancy formation energy is reduced by about 0.9~eV compared with the corresponding DFT value and is in excellent agreement with experiment.  The predicted magnitude of nearest-neighbor relaxations is  about 50\% larger compared to the one obtained within DFT. This remarkable sensitivity to correlation effects is most probably pertinent to other types of defects in iron that are of the crucial importance for mechanical properties and thermodynamics of steels, e.g. interstitial sites, stacking faults and dislocations. 

\begin{acknowledgments}

L.P. acknowledges financial support of the Ministry of Education and Science of the Russian Federation in the framework of Increase Competitiveness Program of NUST MISiS (No. K3-2015-038). L.P.  and S.I.S. acknowledge computational resources provided by the Swedish National Infrastructure for Computing (SNIC) at the National Supercomputer Centre (NSC) and PDC Center for High Performance Computing. S.I.S. acknowledges the Swedish Research Council (VR) Project No. 2014-4750, LiLi-NFM, and the Swedish Government Strategic Research Area in Materials Science on Functional Materials at Link\"{o}ping University (Faculty Grant SFO-Mat-LiU No. 2009 00971). O.P acknowledges support by the FP7/ERC, under Grant Agreement No. 278472-MottMetals. This work was further supported by IDRIS/GENCI Orsay under project t2016091393, and the European Research Council under the European Union’s Seventh Framework Programme (FP7/2007-2013) / ERC Grant Agreements Nr. 617196 (CORRELMAT) and Nr. 319286 (Q-MAC).

\end{acknowledgments}

\bibliography{main}

\clearpage

\section{Supplementary material}

In this supplementary material, we provide additional details concerning
the supercell DFT+DMFT scheme set up for the calculation of the vacancy
formation energies, as well as additional support for the physical picture 
presented in the main text.

\section{Details of the methodology}

Our procedure is to take as a starting point a converged DFT calculation, using the full potential linearized augmented plane-wave (FLAPW) Wien2k code \cite{blaha2001wien2k}, which is then used to perform charge self-consistent DFT+DMFT. We use a k-point mesh with 8x8x8 points for the 16 atom supercell, and 4x4x4 points for the 54 atom supercell. GGA calculations are done with the PBE96 functional. Atomic sphere radii (RMTs) are reduced to 2.12 from the default value of 2.37 to allow for atomic position relaxation. A first set of relaxed atomic positions was obtained in the large supercell by letting the atoms relax in spin-polarized GGA using atomic forces, at the equilibrium volume corresponding to a spin-polarized GGA calculation. The same atomic positions were then used in paramagnetic GGA and DFT+DMFT, at the experimental volume. Then, a second and final set of relaxed positions was obtained by manually relaxing the two first coordination shells, in DFT+DMFT, at the experimental volume. We show that while the GGA relaxed positions are a good starting point, corrections due to many-body effects included within DMFT still modify the nearest-neighbor positions quite significantly.
In order to avoid systematic errors in the vacancy formation energy due to a different computational setup, we used the same cell geometries in calculations with and without the vacancy.

Charge self-consistent DFT+DMFT calculations are performed on the same 2x2x2 and 3x3x3 supercells using the TRIQS library\cite{triqs_main} and its \emph{DFT tools} package\cite{triqs_dft_tools} based on the DFT+DMFT implementation of Ref.\cite{Aichhorn2009} within the Wien2k package \cite{blaha2001wien2k}. We construct Wannier orbitals representing the Fe 3$d$ states from the Kohn-Sham eigenstates within a window [-6.8~eV, 5.4~eV] around the Fermi level using the projection approach of Ref.~\cite{Aichhorn2009}. From the local impurity problem we obtain the DMFT self-energy on the 7 (with vacancy) or 8 (without vacancy) inequivalent iron atoms of our supercell. 
This formidable computational hurdle -- hitherto never overcome -- has been dealt with by resorting to the fast segment representation version of the continuous-time quantum Monte-Carlo  (CTQMC) hybridization expansion algorithm in order to solve the seven impurity problems.


The DFT+DMFT Hamiltonian is 
\begin{equation}
H=H^{LDA} + H^{U} - H^{DC}
\end{equation}

where 

\begin{equation}
H^{U} =  \sum_{mm^\prime,\sigma} U^{\sigma\bar{\sigma}}_{mm^\prime}n_{m\sigma}n_{m^\prime,\bar{\sigma}} + \sum_{m\neq m^\prime,\sigma} U^{\prime\sigma\sigma}_{mm^\prime}n_{m\sigma}n_{m^\prime,\sigma}
\end{equation}

with

\begin{equation}
\begin{split}
U^{\sigma\bar{\sigma}}_{mm^\prime} &= \mathcal{U}_{mm^\prime mm^\prime}  \\
J_{mm^\prime} &= \mathcal{U}_{mm^\prime m^\prime m}  \\
U^{\prime \sigma \sigma}_{mm^\prime} &= U^{\sigma\bar{\sigma}}_{mm^\prime} - J_{mm^\prime}
\end{split}
\end{equation}

Here, $\mathcal{U}$ is the full Slater-parameterized interaction matrix, and $H^U$ is the density-density part of the full Slater Hamiltonian. $H^{DC}$ corrects the double counting of interactions, as both $H^{U}$ and $H^{LDA}$ contain a part of the on-site electron-electron interaction. We take the around mean-field approximation for $H^{DC}$\cite{anisimov_1991,czyzyk_1994}. 

In the calculations for the system with a vacancy, the DMFT 
self-consistency condition corresponds to a set of seven equations,
one for each inequivalent atom $\alpha$. They require the 
impurity Green's functions $G_{\mathrm{imp}}^\alpha(i\omega_{n})$ to equal the 
respective projections of the lattice Green's functions:
\begin{equation} 
\left[G_{\mathrm{imp}}^\alpha(i\omega_{n})\right]_{mm'}=\sum_{\mathbf{k},\nu\nu'} 
P_{\ell m, \nu}^{\alpha}(\mathbf{k})
~G_{\nu\nu'}(\mathbf{k},i\omega_{n})~\left[P_{\ell m',\nu'}^{\alpha}(\mathbf{k})\right]^*
\end{equation}
where the projector $P_{\ell m, \nu}^{\alpha}(\mathbf{k})$ denotes the 
scalar product between the Kohn-Sham state $(\mathbf{k}, \nu)$ and the
local orbital of character $(\ell m)$ on atom $\alpha$. 
The (inverse) correlated Green's function in the Kohn-Sham basis is given by 
\begin{equation} 
 [G^{-1}(\mathbf{k},i\omega_{n})]_{\nu\nu'} = (i\omega_{n} +\mu -\varepsilon_{\mathbf{k}}^{\nu}) \delta_{\nu\nu'} -\Sigma_{\nu\nu'}(\mathbf{k},i\omega_{n}),
\end{equation}
where $\varepsilon_{\mathbf{k}}^{\nu}$ are the 
Kohn-Sham energies and 
$\Sigma_{\nu\nu'}(\mathbf{k},i\omega_{n})$ is the upfolded
self-energy 
\begin{equation} 
\Sigma_{\nu\nu'}(\mathbf{k},i\omega_{n})=
\sum_{\alpha,mm'}
\left[P_{\ell m,\nu}^{\alpha}(\mathbf{k})\right]^*
~\left[\Sigma^{\alpha}(i\omega_{n})\right]_{mm'}
~P_{\ell m',\nu'}^{\alpha}(\mathbf{k}).
\end{equation}
with 
\begin{equation}
 \left[\Sigma^{\alpha}(i\omega_{n})\right]_{mm'} = \left[\Sigma_\mathrm{imp}(i\omega_n)\right]_{mm'}-\left[\Sigma_{DC}\right]_{mm'}
\end{equation}
being the difference between the impurity self-energy
$\Sigma_\mathrm{imp}(i\omega_n)$ and the double-counting correction
$\Sigma_{DC}$.

Finally, the interaction energy is computed as $E_{int} = \frac{1}{2}\sum_{ij}U_{ij}\left<n_in_j\right> - H^{DC}$. The temperature is $\beta=10$, corresponding to $T=1162\text{K}>T_{curie}$, where bcc $\alpha$-Fe is paramagnetic.

About 15 iterations of DFT+DMFT are needed before reaching convergence in the DFT+DMFT cycle. A further averaging of the total energy values is required to obtain precise enough values. In practice, at least 50 more cycles are needed, and the statistical uncertainty shown in table \ref{tab:energies} and Fig. 1 of the main text is the empirical standard deviation of the value over these iterations. This uncertainty increases with the supercell size.

Our $2 \times 2 \times 2$ and $3 \times 3 \times 3$ supercells have an overall cubic symmetry. However,  the presence of a vacancy still breaks the on-site cubic point group symmetries for all iron atoms apart from the central one. For those atoms the Wien2k  code uses   local coordinate frames  chosen in such a way as to have the highest possible on-site symmetry. 
Subsequently, our impurity problems are also solved for those local coordinates. 
Hence,  a corresponding inverse rotation should be applied to the resulting self-energies if one wishes to compare them with that of perfect bcc Fe. The later is, of course, obtained for the standard coordinate frame with $x$, $y$, and $z$ axises along the cube edges.  However, because we are using a density-density Hamiltonian instead of the full rotationally-invariant one, the self-energies obtained for those local frames are still somewhat different from that of perfect bcc iron even after the inverse rotation.  Hence, in our calculations of the ideal supercells we employed the same local coordinate frames as for the supercells with vacancy in order to avoid spurious contributions of those rotations to the vacancy formation energy.  Furthermore, we verified that the off-diagonal elements in the Green's functions stay small, so that we could neglect them.


\section{Vacancy formation energies}

In Table~\ref{tab:energies} we list the vacancy formation energy in bcc Fe obtained by different theoretical approaches and experiments. Figure 2 of the main text corresponds to a graphical representation of these data.

\begin{table}
	\begin{center}
	\resizebox{\linewidth}{!}{
		\begin{tabular}{|l|r|c|}\hline
			Method & $E^f_{vac}$ (eV) & Uncertainty (eV) \\
			\hline
			\multicolumn{3}{|l|}{2x2x2 supercell } \\
			\hline	
			LDA (PM) & 2.50 & $<10^{-2}$ \\
			GGA (PM) & 2.24 & $<10^{-2}$ \\
			GGA (FM) & 2.45 & $<10^{-2}$ \\
			DFT+DMFT & 1.90  & $\pm 0.097$  \\
			\hline
			\multicolumn{3}{|l|}{3x3x3 supercell } \\
			\hline	
			LDA (PM) & 2.33 & $<10^{-2}$ \\
			LDA (PM, relaxed) & unstable &  - \\
			GGA (PM) & 2.06 & $<10^{-2}$ \\
			GGA (PM, relaxed) & unstable &  - \\
			GGA (FM) & 2.71 & $<10^{-2}$ \\
			GGA (FM, relaxed) & 2.50 & $<10^{-2}$ \\
			DFT+DMFT (unrelaxed) & 1.77  & $\pm 0.14$  \\
			DFT+DMFT (FM GGA relaxed) & 1.66  & $\pm 0.15$  \\	
			DFT+DMFT (DMFT relaxed) &   1.56  & $  \pm 0.13 $  \\	
			DFT+DMFT (FM DMFT) &   2.45  & $  \pm 0.15 $  \\	
			\hline
			\multicolumn{3}{|l|}{Experiment (positron annihilation spectroscopy)} \\
			\hline	
			Kim et al. (1978)\cite{kim_positron_1978} & 1.4  & $\pm 0.1$  \\
			Matter et al. (1979)\cite{Matter_positron_1979} &  1.60  &  $\pm$ 0.1 \\
			Schaefer et al. (1977)\cite{schaefer_postitron_1977} & 1.53  & $\pm 0.15$  \\
			Schaefer et al. (1987)\cite{schaefer_positron_1987} & 1.74  & $\pm 0.1$  \\
			de Shepper et al. (1983)\cite{schepper_positron_1983} & 1.79  & $\pm 0.1$  \\
			\hline
			\multicolumn{2}{ | l }{Other DFT refs}  &  \multicolumn{1}{ | c |}{method used} \\
			\hline
			Domain \& Becquart\cite{becquart_iron_2001}   &   1.95   &    GGA PW91  (relaxed, FM)     \\
			S\"oderlind et al.\cite{soderlind_vac_bcc_metal_2000}    &   2.18   &    FP LMTO GGA (relaxed, FM)     \\
			Mizuno et al.\cite{mizuno2001first}     &   2.45    &   FP GGA PW91 (FM)     \\
			Kandaskalov et al.\cite{connetable_multivac_iron_2013}  &  2.16  &   GGA PW91 (relaxed, FM)  \\
			\hline
		\end{tabular}}
		\caption{Vacancy formation energies. The statistical uncertainty shown is the empirical standard deviation of the value over the last 50 iterations. }
		\label{tab:energies}
	\end{center}
\end{table}

\section{Density of states of iron $e_g$ and $t_{2g}$ orbitals}
The density of states of figure 5 in the main part is shown below for the full energy range containing the Fe 3$d$ bands.

\begin{figure}
        \centering
        \includegraphics[width=80mm]{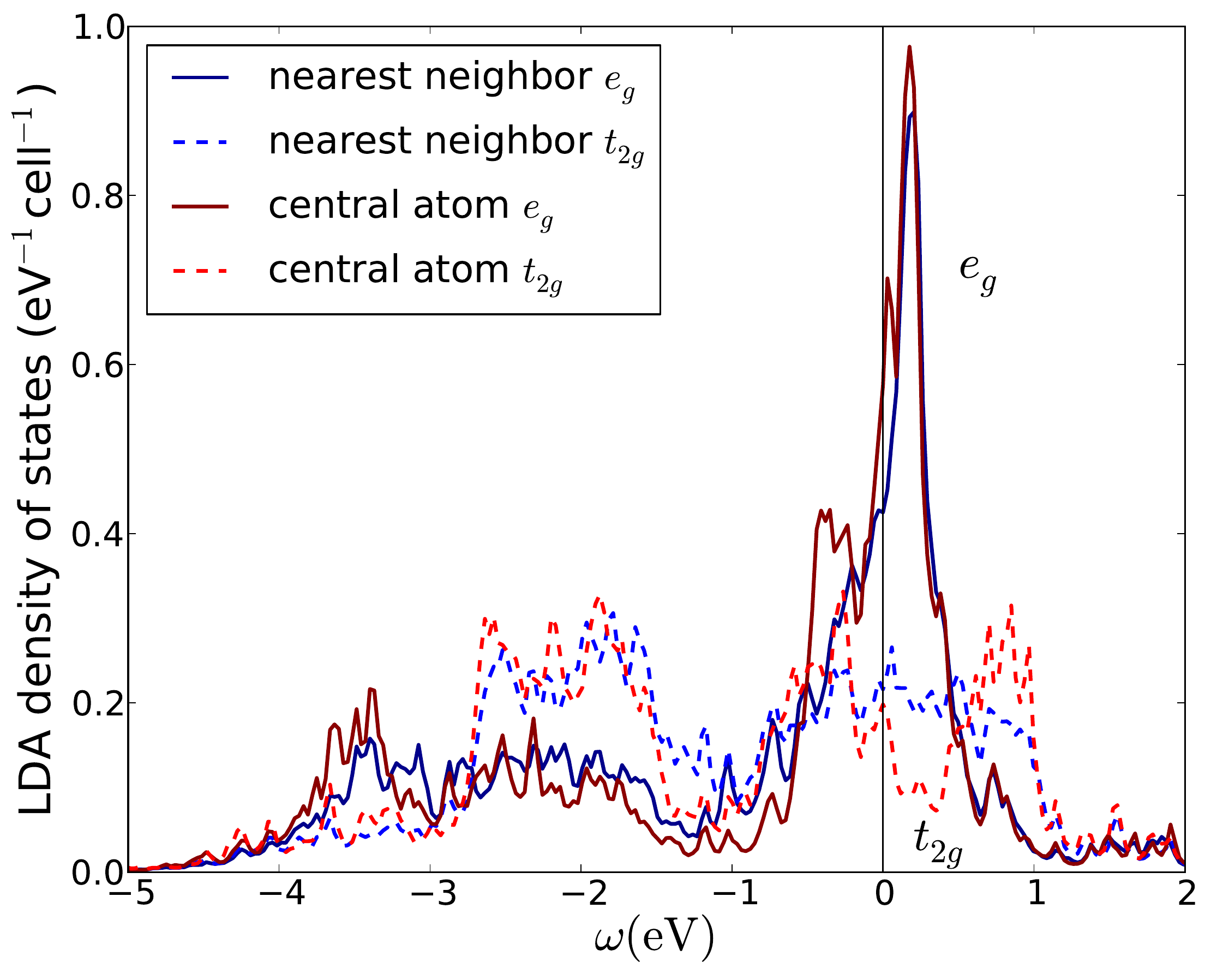}
        \caption{ Density of states on the $e_g$ (full line) and $t_{2g}$ (dashed line)  on the first nearest neighbor to the vacancy (blue) and on the central atom (red). (Color online) }
        \label{fig:full_DOS}
\end{figure}

\bibliography{main}

\end{document}